\documentclass[12pt]{article}
\usepackage[cp1251]{inputenc}
\usepackage[russian]{babel}
\begin{document}
\large
\par
\begin{center}
{\bf Connection Between $\nu_e, \nu_\mu, \nu_\tau$ and $\nu_1,
\nu_2, \nu_3$ Neutrino States and Time Dependence of Neutrino Wave
Functions and Transition Probabilities at Three Neutrino
Oscillations in Vacuum}
\par
\vspace{0.3cm} Beshtoev Kh. M. (beshtoev@cv.jinr.ru)
\par
\vspace{0.3cm} Joint Institute for Nuclear Research, Joliot Curie
6, 141980 Dubna, Moscow region Russia and Scientific Research
Institute of Applied Mathematics and Automation  KBSC of RAS,
Nalchik, Russia.
\end{center}
\vspace{0.3cm}

\par
Abstract

\par
For description of the $d, s, b$ quark mixings the
Cabibbo-Kobayashi-Maskawa matrices are used but they do not
contain the time dependence. In this work the analogous matrix is
obtained for the case of three neutrino ($\nu_{e}, \nu_{\mu },
\nu_{\tau}$) mixings (oscillations) in  vacuum in the general
case, when CP violation is absent. In contrast to the quark case
this matrix contains the time dependence. The matrix for
probability of neutrino transitions (oscillations) in vacuum is
also obtained. Naturally, it contains the time dependence. The
matrix which does not contain the time dependence is obtained by
using time $t$ averaging of this matrix. Elements of this matrix
can be used to describe neutrino decays. \\


\section{Introduction}

At present, existence of three families of leptons and quarks
$$
\begin{array}{cc} u\\ d \end{array};\qquad
\begin{array}{cc}c\\ s \end{array};\qquad
\begin{array}{cc} t \\b \end{array} , \eqno(1)
$$
$$
\begin{array}{cc}  \nu_{e}\\  e \end{array};\qquad
\begin{array}{cc} \nu _{\mu }\\ \mu \end{array};\qquad
\begin{array}{cc}  \nu _{\tau}\\ \tau \end{array}
\eqno(2)
$$
is established [1]. In the framework of the standard model of weak
interactions [2], i.\,e. at $W$, $Z^{\rm 0}$ boson exchanges,
transitions between different families of leptons or quarks with
flavor number violations do not take place. In the quark sector,
mixings between $d, s, b$ quarks (i.\,e. transitions between
different families of quarks) are described by
Cabibbo--Kobayashi--Maskawa matrices $V_q$ [3]
$$
\left(\begin{array}{c} d' \\
s' \\ b' \end{array} \right) =
 V_q \left(\begin{array}{c} d \\
s \\ b \end{array} \right),  \eqno(3)
$$
which are parameterized by 4 parameters, three angle $\theta_q,
\beta_q, \gamma_q$ of rotation and one parameter of CP violation
$\delta'_q$.
$$
{V_q = \left( \begin{array} {ccc}c'_1& s'_1 c'_3 & s'_1 s'_3 \\
-s'_1 c'_2 & c'_1 c'_2 c'_3 - s'_2 s'_3 exp(i \delta'_q) & c'_1
c'_2 s'_3 +
s'_2 s'_3 exp(i \delta'_q) \\
s'_1 s'_2 & -c'_1 s'_2 c'_3 - c'_2 s'_3 exp(i \delta'_q) & -c'_1
s'_2 s'_3 + c'_2 c'_3 exp(i \delta'_q) \end{array} \right)} ,
\eqno(4)
$$
where
$$
c'_1 = \cos {\theta_q } , \quad s'_1 =\sin{\theta_q}, \quad c'^2_1
+ s'^2_1 = 1 ;
$$
$$
c'_2 = \cos {\beta_q }, \quad s'_2 =\sin{\beta_q}, \quad c'^2_2 +
s'^2_2 = 1 ; \eqno(5)
$$
$$
c'_3 = \cos {\gamma_q} , \quad s'_3 =\sin{\gamma_q}, \quad c'^2_3
+ s'^2_3 = 1 ;
$$
$$
\exp(i\delta'_q) = \cos{\delta'_q } + i \sin{\delta'_q} .
$$

\par
It is especially necessary  to remark, that $d', s', b'$ quarks
are superpositions of $d, s, b$ quarks which are eigenstates of
strong interactions.
\par
The charged current in the standard model of weak interactions for
three quark families has the following form [2]:
$$
j^\alpha  = \left(\!\!\begin{array}{ccc}\bar u \bar c \bar t
\end{array}\!\!\right)_L \gamma^\alpha V \left(\begin{array}{c} d \\
s \\ b \end{array} \right)_L,  \eqno(6)
$$
and then the interaction Lagrangian is
$$
{\cal L} = \frac{g}{\sqrt{2}} j^\alpha W^{+}_\alpha  +\rm  h.c.
\eqno(7)
$$
From expression (7) we see that quarks are produced in the weak
interactions in superposition states of $d, s, b$ quark states -
eigenstates of the strong interactions. And there is no time
dependence there. It is necessary to remark that this supposition
requires to be commented. This necessity is related with the fact
that the quarks are produced in the strong interactions and then
for presence of the weak interactions which violate flavor
numbers, they are transformed into superpositions of the strong
interactions eigenstates. The problem is: can the weak
interactions violate aroma numbers at producing quarks by the
strong interactions? If that's the case, then quarks will be
produced in superposition states and there must be no time
dependence otherwise there must be a time dependence [4].
\par
The suggestion that, in analogy with $K^{o},\bar K^{o}$
oscillations, there could be neutrino-antineutrino oscillations (
$\nu \rightarrow \bar \nu$) was considered by Pontecorvo [5] in
1957. It was subsequently considered by Maki et al. [6] and
Pontecorvo [7] that there could be mixings (and oscillations) of
neutrinos of different flavors (i.e., $\nu _{e} \rightarrow \nu
_{\mu }$ transitions).
\par
Below we consider the connection between $\nu_e, \nu_\mu,
\nu_\tau$ and $\nu_1, \nu_2, \nu_3$ neutrino states and time
dependence of neutrino wave functions and transition probabilities
at three neutrino oscillations in vacuum.

\section{Connection Between $\nu_e, \nu_\mu, \nu_\tau$ and $\nu_1,
\nu_2, \nu_3$ Neutrino States and Time Dependence of Neutrino Wave
Functions and Transition Probabilities at Three Neutrino
Oscillations in Vacuum}

At first we consider the connection between $\nu_e, \nu_\mu,
\nu_\tau$ and $\nu_1, \nu_2, \nu_3$ neutrino states at three
neutrino transitions in vacuum and then come to computation of
neutrino wave functions and transition probabilities. The case
when the $CP$ violation is absent will be discussed.

\subsection{Connection Between $\nu_e, \nu_\mu, \nu_\tau$ and
$\nu_1, \nu_2, \nu_3$ Neutrino States at Three Neutrino
Transitions in Vacuum}
\par
We can connect the wave functions of physical neutrino states
$\Psi_{\nu_e}, \Psi_{\nu_\mu}, \Psi_{\nu_\tau}$ with the wave
functions of intermediate neutrino states $\Psi_{\nu_1},
\Psi_{\nu_2}, \Psi_{\nu_3}$ and then
$$
\left(\begin{array}{c} {\nu_e} \\
{\nu_\mu} \\ {\nu_\tau} \end{array} \right) =
V \left(\begin{array}{c} {\nu_1} \\
{\nu_2} \\ {\nu_3} \end{array} \right) \quad \to \eqno(8)
$$
or
$$
\quad \to \quad
\left(\begin{array}{c} \Psi_{\nu_e} \\
\Psi_{\nu_\mu} \\ \Psi_{\nu_\tau} \end{array} \right) =
V \left(\begin{array}{c} \Psi_{\nu_1} \\
\Psi_{\nu_2} \\ \Psi_{\nu_3} \end{array} \right) , \eqno(9)
$$
where the neutrino mixing matrix $V$ can be given [8, 9] in the
following convenient form proposed by Maiani [10] ($\theta =
\theta_{12}, \beta = \theta_{13}, \gamma = \theta_{23}$):
\par
$$
{V {=} \left( \begin{array} {ccc}1& 0 & 0 \\
0 & c_{\gamma} & s_{\gamma} \\ 0 & -s_{\gamma} & c_{\gamma} \\
\end{array} \right)\!\! \left( \begin{array}{ccc} c_{\beta} & 0 &
s_{\beta} \exp(-i\delta) \\ 0 & 1 & 0 \\ -s_{\beta} \exp(i\delta)
& 0 & c_{\beta} \end{array} \right)\!\! \left( \begin{array}{ccc}
c_{\theta} & s_{\theta} & 0 \\ -s_{\theta} & c_{\theta} & 0 \\ 0 &
0 & 1 \end{array}\right)} , \eqno(10)
$$
$$
c_{e \mu} = \cos {\theta } , \quad s_{e \mu} =\sin{\theta}, \quad
c^2_{e \mu} + s^2_{e \mu} = 1 ;
$$
$$
c_{e \tau} = \cos {\beta }, \quad s_{e \tau} =\sin{\beta}, \quad
c^2_{e \tau} + s^2_{e \tau} = 1 ; \eqno(11)
$$
$$
c_{\mu \tau} = \cos {\gamma} , \quad s_{\mu \tau} =\sin{\gamma},
\quad c^2_{\mu \tau} + s^2_{\mu \tau} = 1 ;
$$
$$
 \exp(i\delta) = \cos{\delta } + i \sin{\delta} ,
$$
then $\nu_e, \nu_\mu, \nu_\tau$ neutrino states are transformed
into superpositions of $\nu_1, \nu_2, \nu_3$ neutrino states. And
then the charged current in the standard model of weak
interactions [2] for three lepton families gets the following form
[2]:
$$
j^\alpha  = \left(\!\!\begin{array}{ccc}\bar e \bar \mu \bar \tau
\end{array}\!\!\right)_L \gamma^\alpha V \left(\begin{array}{c} \nu_1 \\
\nu_2 \\ \nu_3 \end{array} \right)_L,  \eqno(12)
$$
and the interaction Lagrangian is
$$
{\cal L} = \frac{g}{\sqrt{2}} j^\alpha W^{+}_\alpha  +\rm  h.c.
\eqno(13)
$$
\par
Using the above matrix $V$, we can write down it in a
component-wise form [8, 9]:
\par
$$
\Psi_{\nu_l} = \sum^{3}_{k=1}V_{\nu_l \nu_k} \Psi_{\nu_{k}},
$$
$$
\Psi_{\nu_k} = \sum^{3}_{k=l}V^{*}_{\nu_k \nu_l} \Psi_{\nu_{l}},
\qquad l = e, \mu, \tau , \qquad k = 1 \div 3 , \eqno(14)
$$
where $\Psi_{\nu_{k}}$ is a wave function of neutrino with
momentum $p$ and mass $m_{k}$. If to separate time dependence of
neutrino wave functions then we get
\par
$$
\Psi_{\nu_{k}}(t) = e^{-i E_k t} \Psi_{\nu_{k}}(0) , \eqno(15)
$$
\par
\noindent
and
$$
\Psi_{\nu _{l}(t)} =\sum^{3}_{k=1} e^{-i E_k t} V_{\nu_l \nu_k}
\Psi_{\nu_{k}}(0)  . \eqno(16)
$$
Using unitarity of matrix $V$ or expression (14) we can rewrite
expression (16) in the following form:
\par
$$
\Psi_{\nu _{l}}(t) = \sum^{}_{l'= e,\mu, \tau} \sum^{3}_{k=1}
V_{\nu_{l'} \nu_k} e^{-i E_k t} V^{*}_{\nu_l \nu_k}
\Psi_{\nu_{l'}(0)} , \eqno(17)
$$
and introducing symbol $b_{\nu_{l}\nu _{l'}}(t)$
$$
b_{\nu_{l}\nu _{l'}}(t) = \sum^{3}_{k=1} V_{\nu_{l'} \nu_k} e^{-i
E_k t} V^{*}_{\nu_l \nu_k} , \eqno(18)
$$
we obtain
$$
\Psi_{\nu_{l}}(t) = \sum^{}_{l'=e, \mu, \tau}
b_{\nu_{l}\nu_{l'}}(t) \Psi_{\nu_{l'}}(0) , \eqno(19)
$$
where $b_{\nu_{l} \nu_{l'}}(t)$-is the amplitude of transition
probability $\Psi_{\nu_{l}}  \rightarrow  \Psi_{\nu_{l'}}$.
\par
\noindent And the corresponding transition probability
$\Psi_{\nu_{l}} \rightarrow \Psi_{\nu_{l'}}$ is:
\par
$$
P_{\nu_{l}\nu_{l'}}(t) =\mid \sum^{3}_{k=1} V^{*}_{\nu_l' \nu_k}
e^{-i E_k t} V_{\nu_l \nu_k} \mid^{2} . \eqno(20)
$$
\noindent It is obvious that
\par
$$
\sum^{}_{l'=e, \mu, \tau} P_{\nu _{l'} \nu_{l}}(t) = 1 . \eqno(21)
$$
\par
Now we will come to computation of neutrino wave functions
$\Psi_{\nu_e}, \Psi_{\nu_\mu}, \Psi_{\nu_\tau}$ and a probability
of transitions (oscillations) of these neutrinos.

\subsection{Expressions for Neutrino Wave Functions and Probability
of $\nu_e, \nu_\mu, \nu_\tau \to \nu_e, \nu_\mu, \nu_\tau$
Transitions (Oscillations) without $CP$ Violation in Vacuum}

If not to take $CP$ violation (i.e., $\delta = 0$) into account
and use expressions (8-14), (14-21), then we obtain the following
expressions for the amplitude of neutrino transitions:
\par
\noindent 1. If primary neutrinos are $\nu_e$ neutrinos, then for
this neutrino wave function for $\nu_e \to \nu_e$, $\nu_e \to
\nu_\mu$, and $\nu_e \to \nu_\tau$ transitions we get
$$
\Psi_{\nu_e \to \nu_e, \nu_\mu, \nu_\tau} (t) = [cos^2 (\beta)
cos^2 (\theta) exp(-i E_1 t) +
$$
$$
+cos^2 (\beta) sin^2 (\theta) exp(-i E_2 t) +
$$
$$
+sin^2 (\beta) exp(-i E_3 t) ] \Psi_{\nu_e}(0) + \eqno(22)
$$
$$
+[cos(\beta) cos(\theta) exp(-i E_1 t)  (-sin(\gamma) sin(\beta)
cos(\theta)-
$$
$$
-cos(\gamma) sin(\theta))+cos(\beta) sin(\theta) exp(-i E_2 t)
(-sin(\gamma) sin(\beta) sin(\theta)+
$$
$$
+cos(\gamma) cos(\theta))+sin(\beta) exp(-i E_3 t)  sin(\gamma)
cos(\beta)] \Psi_{\nu_\mu}(0)
$$
$$
+[cos(\beta) cos(\theta) exp(-i E_1 t)  (-cos(\gamma) sin(\beta)
cos(\theta)+sin(\gamma) sin(\theta))+
$$
$$
+cos(\beta) sin(\theta) exp(-i E_2 t)  (-cos(\gamma) sin(\beta)
sin(\theta)-sin(\gamma) cos(\theta))+
$$
$$
+sin(\beta) exp(-i E_3 t)  cos(\gamma)
cos(\beta)]\Psi_{\nu_\tau}(0) .
$$
\par
Expression (22) can be rewritten in the following form:
$$
\Psi_{\nu_e \to \nu_e, \nu_\mu, \nu_\tau} (t) = b_{\nu_e \nu_e}(t)
\Psi_{\nu_e}(0) + b_{\nu_e \nu_\mu}(t) \Psi_{\nu_\mu}(0) +
b_{\nu_e \nu_\tau}(t) \Psi_{\nu_\tau}(0) , \eqno(22')
$$
where $b_{ ... }$ are coefficients before neutrino wave functions.
\par
\noindent
1.1. Probability of $\nu_e \to \nu_e$ neutrino
transitions obtained from exp. (22) is given by the following
expression:
$$
P_{\nu_e \to \nu_e} (t)= 1 - cos^4(\beta)sin^2(2 \theta) sin^2(- t
(E_1-E_2)/2) -
$$
$$
cos^2(\theta) sin^2(2 \beta) sin^2(- t (E_1-E_3)/2) - \eqno(23)
$$
$$
-sin^2(\theta) sin^2(2 \beta) sin^2(- t (E_2-E_3)/2) .
$$
1.2. Probability of $\nu_e \to \nu_\mu$ neutrino transitions
obtained from exp. (22) is given by the following expression:
$$
P_{\nu_e \to \nu_\mu} (t)=4 cos^2(\beta) cos(\theta) sin(\theta)
[-sin(\gamma) sin(\beta) sin(\theta)+cos(\gamma) cos(\theta)]
$$
$$
[sin(\gamma) sin(\beta) cos(\theta)+cos(\gamma) sin(\theta)]
sin^2(- t (E_1-E_2)/2)-   \eqno(24)
$$
$$
+4 cos^2(\beta) sin(\beta) cos(\theta) sin(\gamma) [sin(\gamma)
sin(\beta) cos(\theta)+cos(\gamma) sin(\theta)]
$$
$$
\cdot sin^2(- t(E_1-E_3)/2)-4 cos^2(\beta) sin(\beta) sin(\theta)
sin(\gamma) [-sin(\gamma) sin(\beta) sin(\theta)+
$$
$$
+cos(\gamma) cos(\theta)] sin^2(- t(E_2-E_3)/2) .
$$
1.3. Probability of $\nu_e \to \nu_\tau$ neutrino transitions
obtained from exp. (22) is given by the following expression:
$$
P_{\nu_e \to \nu_\tau} (t)=4 cos^2(\beta) cos(\theta) sin(\theta)
[-cos(\gamma) sin(\beta) cos(\theta)+ \eqno(25)
$$
$$
+sin(\gamma) sin(\theta)] [cos(\gamma) sin(\beta)
sin(\theta)+sin(\gamma) cos(\theta)] sin^2(- t (E_1-E_2)/2)-
$$
$$
-4 cos^2(\beta) cos(\theta) sin(\beta) cos(\gamma) [-cos(\gamma)
sin(\beta) cos(\theta)+sin(\gamma) sin(\theta)]
$$
$$
\cdot sin^2(- t (E_1-E_3)/2)+4 cos^2(\beta) sin(\theta) sin(\beta)
cos(\gamma) [cos(\gamma) sin(\beta) sin(\theta)+
$$
$$
+sin(\gamma) cos(\theta)] sin^2(- t (E_2-E_3)/2) .
$$
The check has confirmed that $P_{\nu_e \to \nu_e} (t) + P_{\nu_e
\to
\nu_\mu} (t) + P_{\nu_e \to \nu_\tau} (t) = 1.$ \\

\par
\noindent 2. For the case of $\nu_\mu \to \nu_e, \nu_\mu,
\nu_\tau$ transitions we get
$$
\Psi_{\nu_\mu \to \nu_e, \nu_\mu, \nu_\tau} (t) = [cos(\beta)
cos(\theta) exp(-i E_1 t)
$$
$$
(-sin(\gamma) sin(\beta) cos(\theta)-cos(\gamma) sin(\theta))+
$$
$$
+cos(\beta) sin(\theta) exp(-i E_2 t)  (-sin(\gamma) sin(\beta)
sin(\theta)+cos(\gamma) cos(\theta))+
$$
$$
+sin(\beta) exp(-i E_3 t)  sin(\gamma) cos(\beta)] \Psi_{\nu_e}(0)
+
$$
$$
+[(-sin(\gamma) sin(\beta) cos(\theta)-cos(\gamma) sin(\theta))^2
exp(-i E_1 t) +
$$
$$
+(-sin(\gamma) sin(\beta) sin(\theta)+cos(\gamma) cos(\theta))^2
exp(-i E_2 t) +
$$
$$
+sin^2 (\gamma) cos^2 (\beta) exp(-i E_3 t) ] \Psi_{\nu_\mu}(0)+
$$
$$
[(-sin(\gamma) sin(\beta) cos(\theta)- cos(\gamma) sin(\theta))
exp(-i E_1 t)
$$
$$
(-cos(\gamma) sin(\beta) cos(\theta)+sin(\gamma) sin(\theta))+
$$
$$
+(-sin(\gamma) sin(\beta) sin(\theta)+cos(\gamma) cos(\theta))
exp(-i E_2 t)   \eqno(26)
$$
$$
(-cos(\gamma) sin(\beta) sin(\theta)- sin(\gamma)
cos(\theta))+sin(\gamma) cos^2 (\beta)
$$
$$
exp(-i E_3 t) cos(\gamma)] \Psi_{\nu_\tau}(0) .
$$
\par
Expression (26) can be rewritten in the following form:
$$
\Psi_{\nu_\mu \to \nu_e, \nu_\mu, \nu_\tau} (t) = b_{\nu_\mu
\nu_e}(t) \Psi_{\nu_e}(0) + b_{\nu_\mu \nu_\mu}(t)
\Psi_{\nu_\mu}(0) + b_{\nu_\mu \nu_\tau}(t) \Psi_{\nu_\tau}(0) ,
\eqno(26')
$$
where $b_{ ... }$ are coefficients before neutrino wave functions.
\par
2.1. Probability of $\nu_\mu \to \nu_\mu$ neutrino transitions
obtained from exp. (26) is given by the following expression:
$$
P_{\nu_\mu \to \nu_\mu}(t)=1-4 [-sin(\gamma) sin(\beta)
cos(\theta)-cos(\gamma) sin(\theta)]^2
$$
$$
[-sin(\gamma) sin(\beta) sin(\theta)+ cos(\gamma) cos(\theta)]^2
sin^2(-t (E_1-E_2)/2)
$$
$$
-4 [-sin(\gamma) sin(\beta) cos(\theta)-cos(\gamma) sin(\theta)]^2
sin^2(\gamma) cos^2(\beta)
$$
$$
sin^2(-t (E_1-E_3)/2) \eqno(27)
$$
$$
-4 [-sin(\gamma) sin(\beta) sin(\theta)+cos(\gamma) cos(\theta)]^2
sin^2(\gamma)
$$
$$
cos^2(\beta) sin^2(-t (E_2-E_3)/2) .
$$
\par
\noindent 2.2. Probability of $\nu_\mu \to \nu_e$ neutrino
transitions obtained from exp. (26) is given by the following
expression:
$$
P_{\nu_\mu \to \nu_e}(t)=-4 [-sin(\gamma) sin(\beta)
cos(\theta)-cos(\gamma) sin(\theta)] cos^2(\beta) cos(\theta)
$$
$$
[-sin(\gamma) sin(\beta) sin(\theta)+cos(\gamma) cos(\theta)]
sin(\theta) sin^2(-t (E_1-E_2)/2)
$$
$$
-4 [-sin(\gamma) sin(\beta) cos(\theta)-cos(\gamma) sin(\theta)]
cos^2(\beta) \eqno(28)
$$
$$
cos(\theta) sin(\gamma)  sin(\beta) sin^2(-t (E_1-E_3)/2)
$$
$$
-4 [-sin(\gamma) sin(\beta) sin(\theta)+cos(\gamma) cos(\theta)]
cos^2(\beta) sin(\theta) sin(\gamma)  sin(\beta)
$$
$$
sin^2(-t (E_2-E_3)/2)  .
$$
\par
\noindent 2.3. Probability of $\nu_\mu \to \nu_\tau$ neutrino
transitions obtained from exp. (26) is given by the following
expression:
$$
P_{\nu_\mu \to \nu_\tau}(t)=-4 [-sin(\gamma) sin(\beta)
cos(\theta)-cos(\gamma) sin(\theta)]
$$
$$
[-cos(\gamma) sin(\beta) cos(\theta)+sin(\gamma) sin(\theta)]
[-sin(\gamma) sin(\beta) sin(\theta)
$$
$$
+cos(\gamma) cos(\theta)] [-cos(\gamma) sin(\beta)
sin(\theta)-sin(\gamma) cos(\theta)] sin^2(-t (E_1-E_2)/2)
$$
$$
-4 [-sin(\gamma) sin(\beta) cos(\theta)-cos(\gamma) sin(\theta)]
$$
$$
[-cos(\gamma) sin(\beta) cos(\theta)+sin(\gamma) sin(\theta)]
sin(\gamma) cos^2(\beta) cos(\gamma) \eqno(29)
$$
$$
sin^2(-t (E_1-E_3)/2)
$$
$$
-4 [-sin(\gamma) sin(\beta) sin(\theta)+cos(\gamma) cos(\theta)]
$$
$$
[-cos(\gamma) sin(\beta) sin(\theta)-sin(\gamma) cos(\theta)]
sin(\gamma) cos^2(\beta) cos(\gamma)
$$
$$
sin^2(-t (E_2-E_3)/2)  .
$$
The check has confirmed that $P_{\nu_\mu \to \nu_e} (t) +
P_{\nu_\mu \to
\nu_\mu} (t) + P_{\nu_\mu \to \nu_\tau} (t) = 1.$ \\

\par
\noindent 3. For the case of $\nu_\tau \to \nu_e, \nu_\mu,
\nu_\tau$ transitions we get
$$
\Psi_{\nu_\tau \to \nu_e, \nu_\mu, \nu_\tau} (t) = [cos(\beta)
cos(\theta) exp(-i E_1 t)
$$
$$
(-cos(\gamma) sin(\beta) cos(\theta)+sin(\gamma) sin(\theta))+
cos(\beta) sin(\theta) exp(-i E_2 t)
$$
$$
(-cos(\gamma) sin(\beta) sin(\theta)-sin(\gamma) cos(\theta))+
$$
$$
+sin(\beta) exp(-i E_3 t) cos(\gamma) cos(\beta)] \Psi_{\nu_e}(0)
+
$$
$$
+ [(-sin(\gamma) sin(\beta) cos(\theta)-cos(\gamma) sin(\theta))
exp(-i E_1 t)
$$
$$
(-cos(\gamma) sin(\beta) cos(\theta)+ sin(\gamma)
sin(\theta))+(-sin(\gamma) sin(\beta) sin(\theta)+
$$
$$
+cos(\gamma) cos(\theta)) exp(-i E_2 t) (-cos(\gamma) sin(\beta)
sin(\theta)-sin(\gamma) cos(\theta))+
$$
$$
+sin(\gamma) cos^2 (\beta) exp(-i E_3 t) cos(\gamma)]
\Psi_{\nu_\mu}(0)+  \eqno(30)
$$
$$
+ [(-cos(\gamma) sin(\beta) cos(\theta)+sin(\gamma) sin(\theta))^2
exp(-i E_1 t)+
$$
$$
+(-cos(\gamma) sin(\beta) sin(\theta)-sin(\gamma) cos(\theta))^2
exp(-i E_2 t)+
$$
$$
+cos^2 (\gamma) cos^2 (\beta) exp(-i E_3 t)] \Psi_{\nu_\tau}(0) .
$$
\par
Expression (30) can be rewritten in the following form:
$$
\Psi_{\nu_\tau \to \nu_e, \nu_\mu, \nu_\tau} (t) = b_{\nu_\tau
\nu_e}(t) \Psi_{\nu_e}(0) + b_{\nu_\tau \nu_\mu}(t)
\Psi_{\nu_\mu}(0) + b_{\nu_\tau \nu_\tau}(t) \Psi_{\nu_\tau}(0) ,
\eqno(30')
$$
where $b_{ ... }$ are coefficients before neutrino wave functions.
\par
\noindent 3.1. Probability of $\nu_\tau \to \nu_\tau$ neutrino
transitions obtained from exp. (30) is given by the following
expression:
$$
P_{\nu_\tau \to \nu_\tau}(t)=1-4 [-cos(\gamma) sin(\beta)
cos(\theta)+sin(\gamma) sin(\theta)]^2
$$
$$
[-cos(\gamma) sin(\beta) sin(\theta)-sin(\gamma) cos(\theta)]^2
sin^2(-t (E_1-E_2)/2)
$$
$$
-4 [-cos(\gamma) sin(\beta) cos(\theta)+sin(\gamma) sin(\theta)]^2
cos^2(\gamma) cos^2(\beta)
$$
$$
sin^2(-t (E_1-E_3)/2)  \eqno(31)
$$
$$
-4 [-cos(\gamma) sin(\beta) sin(\theta)-sin(\gamma) cos(\theta)]^2
cos^2(\gamma) cos^2(\beta)
$$
$$
sin^2(-t (E_2-E_3)/2)  .
$$
\par
\noindent 3.2. Probability of $\nu_\tau \to \nu_e$ neutrino
transitions obtained from exp. (30) is given by the following
expression:
$$
P_{\nu_\tau \to \nu_e}(t)=-4 [-cos(\gamma) sin(\beta)
cos(\theta)+sin(\gamma) sin(\theta)]
$$
$$
cos^2(\beta) cos(\theta) [-cos(\gamma) sin(\beta)
sin(\theta)-sin(\gamma) cos(\theta)]  sin(\theta)
$$
$$
sin^2(-t (E_1-E_2)/2)
$$
$$
-4 [-cos(\gamma) sin(\beta) cos(\theta)+sin(\gamma) sin(\theta)]
cos^2(\beta) cos(\theta)
$$
$$
cos(\gamma)  sin(\beta) sin^2(-t (E_1-E_3)/2)  \eqno(32)
$$
$$
-4 [-cos(\gamma) sin(\beta) sin(\theta)-sin(\gamma) cos(\theta)]
cos^2(\beta) sin(\theta) cos(\gamma) sin(\beta)
$$
$$
sin^2(-t (E_2-E_3)/2)  .
$$
\par
\noindent 3.3. Probability of $\nu_\tau \to \nu_\mu$ neutrino
transitions obtained from exp. (30) is given by the following
expression:
$$
P_{\nu_\tau \to \nu_\mu}(t)=-4 [-cos(\gamma) sin(\beta)
cos(\theta)+sin(\gamma) sin(\theta)]
$$
$$
[-sin(\gamma) sin(\beta) cos(\theta)-cos(\gamma) sin(\theta)]
$$
$$
[-cos(\gamma) sin(\beta) sin(\theta) -sin(\gamma) cos(\theta)]
\eqno(33)
$$
$$
[-sin(\gamma) sin(\beta) sin(\theta)+cos(\gamma) cos(\theta)]
sin^2(-t (E_1-E_2)/2)
$$
$$
-4 [-cos(\gamma) sin(\beta) cos(\theta)+sin(\gamma) sin(\theta)]
$$
$$
[-sin(\gamma) sin(\beta) cos(\theta)-cos(\gamma) sin(\theta)]
cos(\gamma) cos^2(\beta) sin(\gamma)
$$
$$
sin^2(-t (E_1-E_3)/2)
$$
$$
-4 [-cos(\gamma) sin(\beta) sin(\theta)-sin(\gamma) cos(\theta)]
$$
$$
[-sin(\gamma) sin(\beta) sin(\theta)+cos(\gamma) cos(\theta)]
$$
$$
cos(\gamma) cos^2(\beta) sin(\gamma) sin^2(-t (E_2-E_3)/2)  .
$$
The check has confirmed that $P_{\nu_\tau \to \nu_e} (t) +
P_{\nu_\tau \to
\nu_\mu} (t) + P_{\nu_\tau \to \nu_\tau} (t) = 1.$ \\
\par
We can rewrite expressions (22'), (26'), (30') for three neutrinos
wave functions in the following compact form:
$$
\left ( \begin{array}{c} \Psi_{\nu_e \to \nu_e, \nu_\mu, \nu_\tau} (t)\\
\Psi_{\nu_\mu \to \nu_e, \nu_\mu, \nu_\tau} (t)
\\ \Psi_{\nu_\tau \to \nu_e, \nu_\mu, \nu_\tau} (t)\\ \end{array} \right) =
\left( \begin{array}{ccc} b_{\nu_e \nu_e}(t)& b_{\nu_e \nu_\mu}(t)&
b_{\nu_e \nu_\tau}(t)\\
b_{\nu_\mu \nu_e}(t)& b_{\nu_\mu \nu_\mu}(t)& b_{\nu_\mu \nu_\tau}(t)\\
b_{\nu_\tau \nu_e}(t)& b_{\nu_\tau \nu_\mu}(t)& b_{\nu_\tau
\nu_\tau}(t)
\end{array} \right)
\left ( \begin{array}{c} \Psi_{\nu_e}(0)\\ \Psi_{\nu_\mu}(0) \\
\Psi_{\nu_\tau}(0)\\ \end{array} \right) . \eqno(34)
$$
\par
Usually matrix $V$ (8) which connects the $\nu_e, \nu_\mu,
\nu_\tau$ neutrino states with  the $\nu_1, \nu_2, \nu_3$ is named
Cabibbo-Kobayashi-Maskawa mixing matrix type for neutrinos. It is
necessary to remark that in reality this matrix is matrix $V(t)$
from expression (34)
$$
\left ( \begin{array}{c} \Psi_{\nu_e \to \nu_e, \nu_\mu, \nu_\tau} (t)\\
\Psi_{\nu_\mu \to \nu_e, \nu_\mu, \nu_\tau} (t)
\\ \Psi_{\nu_\tau \to \nu_e, \nu_\mu, \nu_\tau} (t)\\
\end{array} \right) = V(t)
\left ( \begin{array}{c} \Psi_{\nu_e}(0)\\ \Psi_{\nu_\mu}(0) \\
\Psi_{\nu_\tau}(0)\\ \end{array} \right) , \eqno(35)
$$
where
$$
V(t) = \left( \begin{array}{ccc} b_{\nu_e \nu_e}(t)& b_{\nu_e
\nu_\mu}(t)&
b_{\nu_e \nu_\tau}(t)\\
b_{\nu_\mu \nu_e}(t)& b_{\nu_\mu \nu_\mu}(t)& b_{\nu_\mu \nu_\tau}(t)\\
b_{\nu_\tau \nu_e}(t)& b_{\nu_\tau \nu_\mu}(t)& b_{\nu_\tau
\nu_\tau}(t)
\end{array} \right) . \eqno(36)
$$
We see that in the case of neutrinos mixing matrix $V(t)$ is a
function of time $t$. The analogous expression (3) for quarks
mixing matrix $V_q$ does not contain the time dependence. And at
$t = 0$ this mixing matrix, in contrast to the quark case (see
(30)), has the following diagonal form:
$$
V(t = 0) = \left( \begin{array}{ccc} 1 & 0 & 0 \\
0 & 1 & 0 \\
0 & 0 & 1 \end{array} \right) . \eqno(37)
$$
\par
We can also introduce matrix $V_{prob}(t)$ probabilities of three
neutrino transitions (oscillations) in dependence on time and
write three neutrino transition probabilities in the following
compact form:
$$
V_{prob}(t) = \left( \begin{array}{ccc} P_{\nu_e \to \nu_e} (t)
& P_{\nu_e \to \nu_\mu} (t)& P_{\nu_e \to \nu_\tau} (t)\\
P_{\nu_\mu \to \nu_e} (t)& P_{\nu_\mu \to \nu_\mu} (t)& P_{\nu_\mu
\to \nu_\tau} (t)\\
P_{\nu_\tau \to \nu_e} (t)& P_{\nu_\tau \to \nu_\mu} (t)&
P_{\nu_\tau \to \nu_\tau} (t)
\end{array} \right) . \eqno(38)
$$
If to average $V(t)$  on time $t$ in expression (38), then we
obtain matrix
$$
V = \overline {V (t)} ,  \eqno(39)
$$
which has no time dependence. For this purpose it is necessary in
expressions (23-25), (27-29), (31-33) for $P_{\nu_l \to \nu_{l'}}
(t)$ fulfill the following replacements:
$$
\overline {sin^2(-t (E_1-E_2)/2)} = \frac{1}{2},
$$
$$
\overline {sin^2(-t (E_1-E_3)/2)} = \frac{1}{2}, \eqno(40)
$$
$$
\overline {sin^2(-t (E_1-E_3)/2)} = \frac{1}{2}.
$$
Elements of matrix (39) can be used for description of neutrino
decays.

\section{Conclusion}
\par
The Cabibbo-Kobayashi-Maskawa matrices [3] are used to describe
the $d, s, b$ quark mixings and these matrices do not contain the
time dependence. In this work the analogous matrix is obtained for
the case of three neutrino ($\nu_{e}, \nu_{\mu }, \nu_{\tau}$)
mixings (oscillations) in  vacuum in the general case, when CP
violation is absent. In contrast to the quark case this matrix
contains the time dependence. The matrix for probability of
neutrino transitions (oscillations) in vacuum is also obtained.
Naturally, it contains the time dependence. The matrix which does
not contain the time dependence is obtained by using the time $t$
averaging of this matrix. Elements of this matrix can be used
to describe neutrino decays. \\

\par
{\bf References}\\
\par
\noindent 1. Review of Part. Prop., Phys. Rev. D. 2002. V.66, No.
1. P. 010001.
\par
\noindent 2. S. L. Glashow, Nucl. Phys. 1961. V. 22. P. 579;
\par
S. Weinberg, Phys. Rev. Lett. 1967. V. 19, P. 1264;
\par
A. Salam, Proc. of the 8th Nobel Symp., Ed. N. Svarthholm.
Almgvist
\par
and Wiksell, Stockholm, 1968. P. 367.
\par
\noindent 3. N. Cabibbo, Phys. Rev. Lett. 1963. V. 10. P. 531;
\par
M. Kobayashi, K. Maskawa K., Prog. Theor. Phys. 1973. V. 49. P.
652.
\par
\noindent 4. Kh. M. Beshtoev, JINR Communication E2-2004-182,
\par
Dubna, 2004; hep-ph/0503157;

\par
\noindent 5. B. M. Pontecorvo , Soviet Journ. JETP, 33 (1957) 549;
\par
JETP, 34 (1958) 247.
\par
\noindent 6. Z. Maki et al., Prog.Theor. Phys., 28 (1962) 870.
\par
\noindent 7. B. M. Pontecorvo, Soviet Journ. JETP, 53 (1967) 1717.
\par
\noindent 8. S.M. Bilenky, B.M. Pontecorvo, Phys. Rep., C41 (1978)
225;
\par
F. Boehm, P. Vogel, Physics of Massive Neutrinos: Cambridge
\par
Univ. Press, 1987, p.27, p.121;
\par
S. M. Bilenky, S. P. Petcov, Rev. of Mod.  Phys., 59 (1977) 631.
\par
 V. Gribov, B. M. Pontecorvo, Phys. Lett. B, 28 (1969) 493.
\par
\noindent 9. Kh. M. Beshtoev, JINR Communication E2-2004-58,
Dubna, 2004;
\par
hep-ph/0406124; JINR Communication E2-2005-123, Dubna, 2005,
\par
hep-ph/0506248.
\par
\noindent 10. L. Maiani, Proc. Intern. Symp.  on Lepton--Photon
Interaction,
\par
DESY, Hamburg. P.867, 1977.

\end{document}